\documentclass[symmetry,article,accept,moreauthors,pdftex,10pt,a4paper]{Definitions/mdpi} 

\usepackage{booktabs} 
\usepackage{multirow}
\usepackage{soul} 
\usepackage{microtype}
\usepackage{graphicx}
\usepackage{slashed}
\usepackage{physics}
\usepackage{float}
\setitemize{parsep=6pt,itemsep=0pt,leftmargin=*,labelsep=5.5mm}
\setenumerate{parsep=6pt,itemsep=0pt,leftmargin=*,labelsep=5.5mm}
\setlist[description]{itemsep=0mm}   
\usepackage{units}
\usepackage{textcomp}
\usepackage{amsmath}
\usepackage{amsthm}
\usepackage{graphicx}
\usepackage{esint}
\usepackage{braket}
\firstpage{1} 
\pubvolume{00}
\issuenum{0}
\articlenumber{0}
\pubyear{2019}
\copyrightyear{2019}
\history{Received: 27 February 2019; Accepted: 30 March 2019; Published: date}
\issuenum{0} 
\updates{yes}
 
\pdfoutput=1

\Title{Solving the Gap Equation of the NJL Model through Iterations: Unexpected Chaos}

\Author{Angelo Mart\'inez \orcidA{} and Alfredo Raya \orcidB{}}

\AuthorNames{Angelo Martínez and Alfredo Raya}

\address[1]{%
Instituto de F\'{\i}sica y Matem\'aticas, Universidad Michoacana de San Nicol\'as de Hidalgo, Edificio C-3, \newline Ciudad Universitaria, Francisco J. M\'ugica s/n, Col. Fel\'{\i}citas del R\'{\i}o, Morelia 58040, Michoacan, Mexico. \iffalse; raya@ifm.umich.mx\fi}

\corres{\hangafter=1 \hangindent=1.05em \hspace{-0.82em}Correspondence: angelo@ifm.umich.mx; raya@ifm.umich.mx}

\abstract{We explore the behavior of the iterative procedure to obtain the solution to the gap equation of the Nambu--Jona-Lasinio (NLJ) model  for arbitrarily large values of the coupling constant and in the presence of a magnetic field and a thermal bath. We find that the iterative procedure shows a~different behavior depending on the regularization scheme used. It is stable and very accurate when a hard cut-off is employed. Nevertheless, for the Paul-Villars and proper time regularization schemes,  there exists a value of the coupling constant (different in each case) from where the procedure becomes chaotic and does not converge any longer.}

\keyword{Nambu--Jona-Lasinio model; superstrong coupling; magnetic field; heat bath}

\begin{document}


\section{Introduction}\label{sec1}

The Nambu--Jona-Lasinio (NJL) model represents one of the earliest attempts to describe the strong interactions between protons and neutrons inside an atomic nucleus. Even though today we know a better suited description of these interactions at the fundamental level, namely, among quarks, in the form of the gauge theory called Quantum Chromodynamics (QCD), the simplicity of the NJL model and its ability to gasp, at least to some extent,  robust details of the strong interactions make it a~favorite  candidate to explore the dynamic of strongly interacting particles under extreme conditions. The NJL model describes Dirac fermions---quarks in modern considerations---that interact at contact in the form of a four-fermion interaction  with a single  coupling constant. One of the main features of the model, and the one we are interested in, is the phenomenon of spontaneous chiral symmetry breaking through an analogy to the BCS theory of superconductivity \cite{NJL1961, NJL1961a}.  As an outcome of this symmetry breaking, the fundamental fermion fields acquire a mass that turns out to be a constant.  Thus, to~give this prediction a physically reliable character, it should be valid only up to some scale. In this sense, the model is non-renormalizable, and thus for the loop integrals to converge, the introduction of a~regularization scheme is necessary. The model has been extended to provide it with more reliable features, close to those exhibited in QCD. It can be coupled to the Polyakov loop through an effective potential to describe confinement \cite{Fukushima:2008wg}. It can also be regularized non-locally such that the quark mass function smoothly varies with energy \cite{GomezDumm:2006vz} . Other variants make the model suitable for the description of strong interactions in an effective and refined manner. For a review on the NJL model, we refer the readers to the seminal works in~\cite{Klevansky:1992qe,Buballa:2003qv}. 

On the other hand, addition of a heat bath or density effects in the model is straightforward. These ingredients are the key to study the QCD phase diagram to understand the last transition that hadronic matter experienced during the early Universe as it expanded and cooled down.  Moreover, the recent interest of incorporating the effect of an external magnetic field has been boosted by the fact that in heavy-ion collisions, very strong magnetic fields are generated in the interaction region, changing the behavior of the above transition \cite{Endrodi:2015oba,PhysRevD.86.071502}. Under these circumstances, the role of these external agents in the NJL model is translated into an effective enhancement or dilution of the strength of the coupling constant as compared with that of vacuum~\cite{Shovkovy2013,PhysRevD.86.071502}. Therefore, it is worth exploring the parametric behavior of the solution to the gap equation of the model for arbitrarily large values of the coupling. This is the main subject of this article.

A favorite strategy to solve the gap equation is to cast it in a transcendental form and then use iterations to find the dynamical mass self-consistently~\cite{Aoki:2013dia,Aoki:2019ygj}. Each iteration is then explained in terms of the number of self-energy insertions into the fermion propagator that are being re-summed non-perturbatively. The spirit of this approach is that starting with a given initial guess, after a finite number of iteration steps, one is arbitrarily close to the actual solution. This seemingly harmless procedure has in fact proven to be very robust for the hard cut-off regulators even for large values of the coupling constant. However, for other regulation schemes such as the Pauli-Villars and proper time, surprises appear from the behavior of this procedure, as was noticed by \cite{Aoki:2013dia,Aoki:2019ygj, Martinez:2017wdu} and we revise in deeper detail  in this work. To this end, we have organized the remaining of the article as follows: In Section~\ref{sec:njl} we review the basics of the model. In Section~\ref{sec:gapvac} we present the parametric form of the gap equation for the different regularization schemes, whereas Section~\ref{sec:solvac} is devoted to describing the iterative procedure to find the solution of the gap equation in vacuum for different regularization schemes. We add the influence of a magnetized, thermal medium and explore once more the traits of the iterative solution in Section~\ref{sec:medium}. Conclusion and final remarks are presented in Section~\ref{sec:final}.
 
\section{NJL Model and the Gap Equation}\label{sec:njl}

The Lagrangian for the NJL model is
\begin{linenomath*}
\begin{equation}
\mathcal{L}=\overline{\psi}(i\slashed{\partial}-m_{q})\psi+G\left\{ \left(\overline{\psi}\psi\right)^{2}+\left(\overline{\psi}i\gamma^{5}\vec{\tau}\psi\right)^{2}\right\}\label{njl}.   
\end{equation}
\end{linenomath*}

In modern considerations, $\psi$ is regarded the quark field, $m_q$ is the current quarks mass, $G$ is the coupling constant and $\vec{\tau}$ are the Pauli matrices acting on isospin space. In this Lagrangian, only scalar and pseudoscalar interactions are considered. To explicitly observe the spontaneous breaking of chiral symmetry in the model, we commence from the gap equation. Such an equation states important features of general Quantum Field Theories with spontaneous symmetry breaking, among which it highlights the fact that the ground state does not possess the same symmetry of the underlying Lagrangian. Breaking of the chiral symmetry of the model is due to an interaction which induces a pairing of particles and anti-particles (chiral condensate) resembling the Cooper pair structure of a Type-II Superconductor. Even from the early discussion of the model, within the  Hartree-Fock approximation, the gap equation emerges as a constraint that minimizes the total energy of the system. In the local version of the model, it  has the simple form 
\begin{linenomath*}
\begin{equation}
m=m_{q}-2G\left\langle \bar{\psi}\psi\right\rangle,
\label{gapnjl}    
\end{equation}
\end{linenomath*}
where $m$ is the dynamically generated mass and $-\left\langle \bar{\psi}\psi\right\rangle$ is the so-called chiral condensate, defined as 
\begin{linenomath*}
\begin{equation}
-\left\langle \bar{\psi}\psi\right\rangle =\int\frac{d^{4}k}{(2\pi)^{4}}{\rm Tr}[iS(k)], 
\label{condensate}
\end{equation}
with $S(k)$ the dressed quark propagator 
\begin{equation}
S(k)=\frac{\left(\slashed k+m\right)}{k^{2}-m^{2}}.
\label{prop}
\end{equation}
\end{linenomath*}

One of the characteristics of the NJL model is that the predicted dynamically generated mass is independent of the momentum, making the model an effective description of low-energy QCD. As~we mentioned before, the model is non-renormalizable, and thus a regularization scheme must be used for the integrals involved to converge. For the purposes of this work, we review some of the most commonly used schemes, namely, the $3D$-cut-off ($3D$), the $4D$-cut-off ($4D$), Pauli-Villars (PV) and proper time (PT) regularizations.  Then, the gap equation in  vacuum in each of these regularization schemes is written as
\begin{linenomath*}
\begin{eqnarray}
m&=&\frac{GN_{f}N_{c}}{\pi^{2}}m\left[\Lambda_{3D}\sqrt{\Lambda_{3D}^{2}+m^{2}}-m^{2}\ln\left(\frac{\Lambda_{3D}+\sqrt{\Lambda_{3D}^{2}+m^{2}}}{m}\right)\right],
\label{3D}\\
m&=&\frac{GN_{f}N_{c}}{2\pi^{2}}m\left[\Lambda_{4D}^{2}-m^{2}\ln\left(\frac{\Lambda_{4D}^{2}+m^{2}}{m^{2}}\right)\right],
\label{4D}\\
m&=&\frac{GN_{f}N_{c}}{2\pi^{2}}m\left[\Lambda_{PV}^{2}-m^{2}+m^{2}\ln\left[\frac{m^{2}}{\Lambda_{PV}^{2}}\right]\right],
\label{PV}\\
m&=&\frac{GN_{f}N_{c}}{2\pi^{2}}m\intop_{1/\Lambda_{PT}^2}^{\infty}ds\frac{e^{-m^{2}s}}{s^{2}} =\frac{GN_{f}N_{c}}{2\pi^{2}}m\left[\Lambda_{PT}^2e^{-m^{2}/\Lambda_{PT}^2}-m^{2}\Gamma\left(0,\frac{m^{2}}{\Lambda_{PT}^2}\right)\right],
\label{TP}
\end{eqnarray}
\end{linenomath*}
respectively. Here, $N_f$ is the number of quark families, $N_c$ is the number of colors considered and $\Gamma(a,z)$ is the incomplete gamma function defined as
\begin{linenomath*}
\begin{equation}
\Gamma(a,z)=\int_{z}^{\infty}t^{a-1}e^{-t}dt.
\end{equation}
\end{linenomath*}

To ease the analysis of  the solutions to Equations~(\ref{3D})--(\ref{TP}), 
we express the above relation  in  dimensionless form, 
\begin{linenomath*}
\begin{eqnarray}
    M&=&G'M\left\{ \sqrt{1+M^{2}}-M^{2}\ln\left[\frac{1+\sqrt{1+M^{2}}}{M}\right]\right\} ,\qquad(3D\,\textrm{cut-off})\label{A3D}\\
M&=&G'M\left\{ 1-M^{2}\ln\left[1+M^{-2}\right]\right\} ,\qquad(4D\,\textrm{cut-off })\label{A4D}\\
M&=&G'M\left\{ 1-M^{2}+M^{2}\ln\left[M^{2}\right]\right\} ,\qquad(\textrm{Pauli-Villars)}\label{APV}\\
M&=&G'M\int_{1}^{\infty}d\tau\frac{e^{-M^{2}\tau}}{\tau^{2}} =G'M\left\{e^{-M^{2}}-M^{2}\Gamma(0,M^{2})\right\}, \qquad(\textrm{Proper Time)}\label{ATP}
\end{eqnarray}
\end{linenomath*}
\textls[-20]{where we introduce the dimensionless quantities $M=m/\Lambda$ for all regularizations and  $G'=GN_fN_c/(2\pi^2)$ for 4D,} PV, and PT, whereas $G'=GN_fN_c/\pi^2$ for the 3D regularization schemes.
Below we introduce the method to solve the gap equations in Equations~(\ref{A3D})--(\ref{ATP}) using iterations.

\section{Solving the Gap Equation with Iterations}\label{sec:gapvac}

\begin{figure}[H]
\centering
\includegraphics[width=0.74\textwidth]{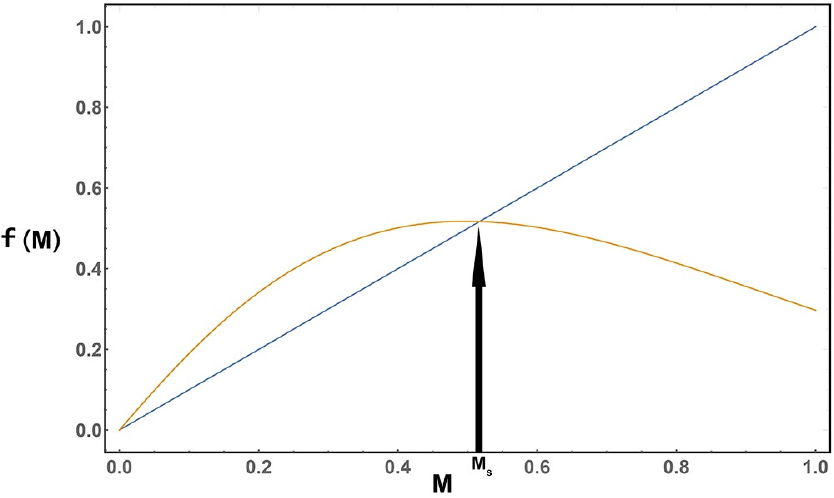}
\caption{Plot of the system of Equations~(\ref{SEq}), were we have used  the RHS of Equation~(\ref{ATP}) in place of $G' f(M)$. The solution of the gap equation is the intersection of the two curves, $M_s$.}
\label{fig:sistemaeq}
\end{figure}

To solve the gap equations described above, we introduce the notation 
\begin{equation}
\begin{array}{rcl}
y&=&M,\\
\hat{y}&=&G'f(M),
\label{SEq}
\end{array}
\end{equation}
where $f(M)$ is the function on the right-hand-side (RHS) of any of the gap equations in Equations~(\ref{A3D})--(\ref{ATP}). If we plot the two equations $y$ and $\hat y$ as  functions of $M$ for a fixed value of $G'=2$, for example, we get Figure~\ref{fig:sistemaeq}.

As we can see from Figure~\ref{fig:sistemaeq}, the solution of the system of equations in Equation~(\ref{SEq}) is the intersection of the two curves. Our goal is then to catch this solution using iterations. The iterating procedure can be depicted in a cobweb plot, a visual tool that depicts the iterative process by joining with straight lines the beginning (vertical) and end (horizontal) of each iteration, Figure~\ref{fig:cobwebej}. For each value of the coupling constant, we can construct the corresponding  cobweb plot. Our main interest is to locate  the point where the iterations converge. By definition, this point corresponds to the solution of the gap equation for that particular value of $G'$. We thus apply the following iteration procedure to the Equations~(\ref{A3D})--(\ref{ATP}):

\begin{figure}[H]
\centering
\includegraphics[width=0.74\textwidth]{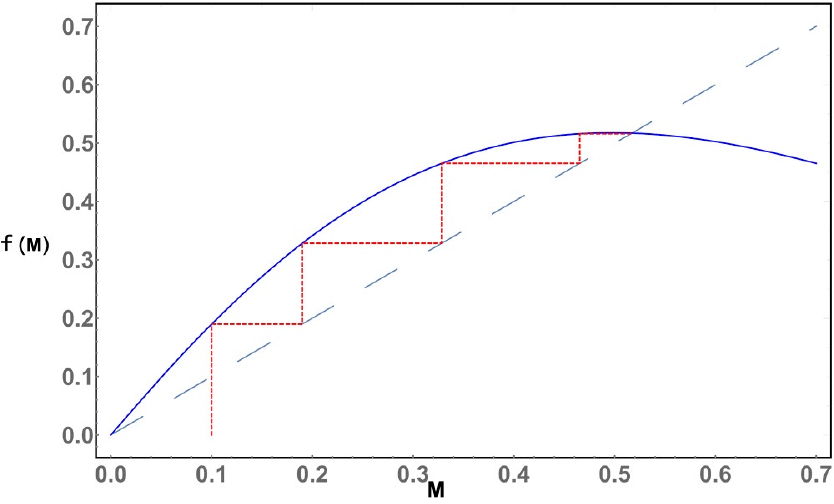}
\caption{Cobweb plot of the  system of Equations~(\ref{SEq}) using Equation~(\ref{ATP}) as the RHS with $G'=2$. Given the initial condition $M=0.1$, a vertical line is drawn till it hits the curve of $\hat{y}$ and from there, a~horizontal line is drawn to the curve of $y$, giving the value of $M$  for the first iteration, and the process in then repeated with this value as the new initial guess. After a few iterations, the procedure converges to the solution of the system of equations.}
\label{fig:cobwebej}
\end{figure}

As we can see from Figure~\ref{fig:sistemaeq}, the solution of the system of equations in Equation~(\ref{SEq}) is the intersection of the two curves. Our goal is then to catch this solution using iterations. The iterating procedure can be depicted in a cobweb plot, a visual tool that depicts the iterative process by joining with straight lines the beginning (vertical) and end (horizontal) of each iteration, Figure~\ref{fig:cobwebej}. For each value of the coupling constant, we can construct the corresponding  cobweb plot. Our main interest is to locate  the point where the iterations converge. By definition, this point corresponds to the solution of the gap equation for that particular value of $G'$. We thus apply the following iteration procedure to the Equations~(\ref{A3D})--(\ref{ATP}):

\begin{itemize}[leftmargin=*,labelsep=5.8mm]
\item	We select a value of $G_0'$ to start with,
\item	We pick a positive initial random value of mass and plug it in the RHS of the gap equation,
\item	For the selected value of the coupling, $G_0'$, we perform a sufficiently large number of iterations; we use 500 iterations,
\item	We store and plot the last 70 iterations obtained; for a converging solution, all these 70 values should be on top of each other in the corresponding plot.
\item   We then pick another value of $G'$ and repeat.
\end{itemize}

In the next section we show the results obtained from this iteration procedure. We insist that this procedure corresponds to self-energy insertions into the fermion propagator in a non-perturbative re-summation~\cite{Aoki:2013dia,Aoki:2019ygj}.

\section{Results: Gap Equation in Vacuum}\label{sec:solvac}

\begin{figure}[H]
\centering
\includegraphics[width=0.7\textwidth]{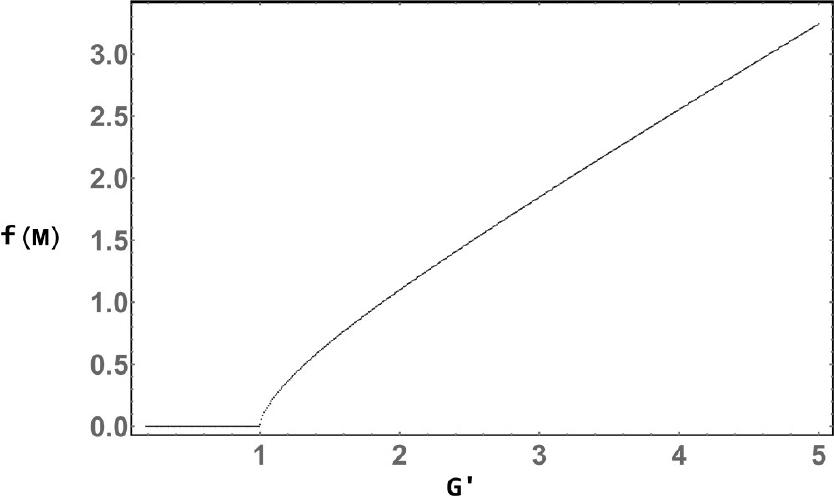}
\caption{Plot of the last 70 iterations to achieve the solution to the gap Equation~(\ref{A3D}). We can observe the critical value of the coupling $G'_c=1$ from where chiral symmetry is broken.}
\label{fig:iteracion3D}
\end{figure}

Now we proceed to analyze the results given by the iterative method described above. In Figure~\ref{fig:iteracion3D}, we plot the last 70 steps of the iteration procedure carried out for Equation~(\ref{A3D}) for each value of the coupling $G'$. As we can see, the plotted curve gives  the characteristic behavior expected for a solution of the gap equation; for a range of values for the coupling, where it is not strong enough, the solution to the gap equation is just the trivial one $M=0$. Then, at the critical value of the coupling $G'_c=1$,  masses start to be dynamically generated. It should be noted nevertheless that because we are working with dimensionless quantities,  the value of $G_c$  changes for different regularization schemes. For example, in Equation~(\ref{3D}), $G_c=\nicefrac{\pi^2}{6}$, provided we take $N_f=2$ and $N_c=3$. Finally, we observe a monotonic growth of the dynamical mass as a function of the coupling $G'$.

\begin{figure}[H]
\centering
\includegraphics[width=0.71\textwidth]{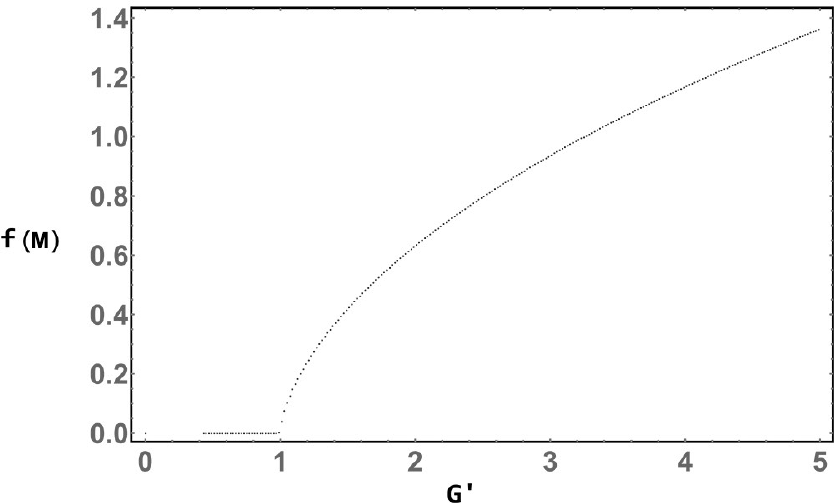}
\caption{Plot of the last 70 iterations to achieve the solution to the gap Equation~(\ref{A4D}). As in Figure~\ref{fig:iteracion3D}, the critical value of the coupling is $G'_c=1$; this is due to the use of dimensionless quantities in the gap~equation.}
\label{fig:iteracion4D}
\end{figure}

Next, for Equation~(\ref{A4D}), the iterative procedure is shown in Figure~\ref{fig:iteracion4D}. Just like in the previous case, there is a critical value for the coupling, $G'_c=1$, to break the chiral symmetry. In this regularization scheme, we can see that the mass grows monotonically, but slowly as a function of the coupling constant in comparison with the $3D$ regularization scheme.

\begin{figure}[H]
\centering
\includegraphics[width=0.71\textwidth]{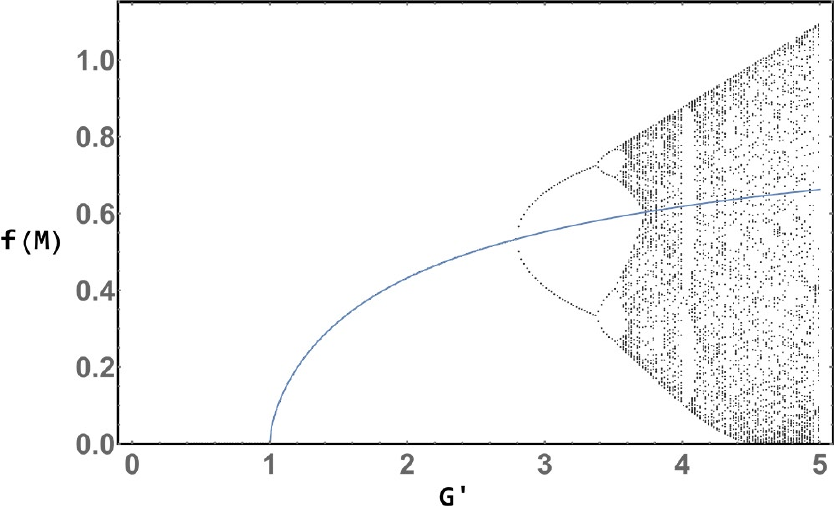}
\caption{Plot of the iterative procedure to solve the gap equation using Equation~(\ref{APV}). As we observed before, if the coupling strength exceeds  $G'_c=1$, chiral symmetry is broken. On the other hand, if the coupling exceeds $G'\approx2.8$,  the iteration procedure ceases to converge, and iterations bifurcate into two branches. As the coupling increases even further, the iterations become chaotic. The solid (blue) curve represent the actual solution to the gap Equation ~(\ref{APV}) for superstrong coupling.}
\label{fig:iteracionPV}
\end{figure}

\begin{figure}[H]
\centering
\begin{tabular}{c}
\includegraphics[width=0.7\linewidth]{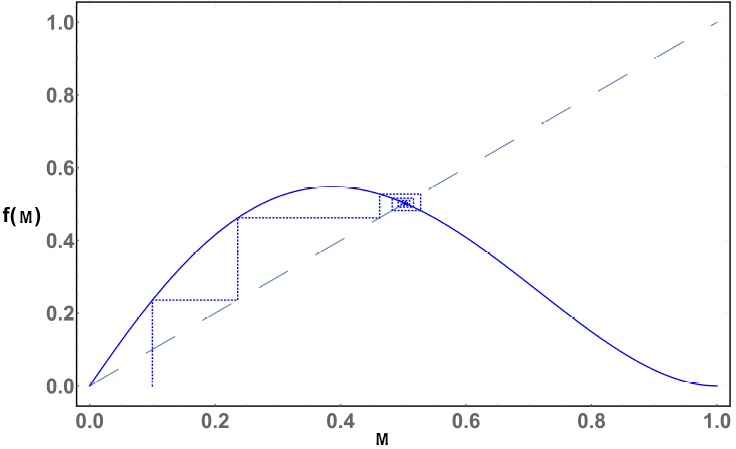} \\
({\bf a}) \\
\includegraphics[width=0.7\linewidth]{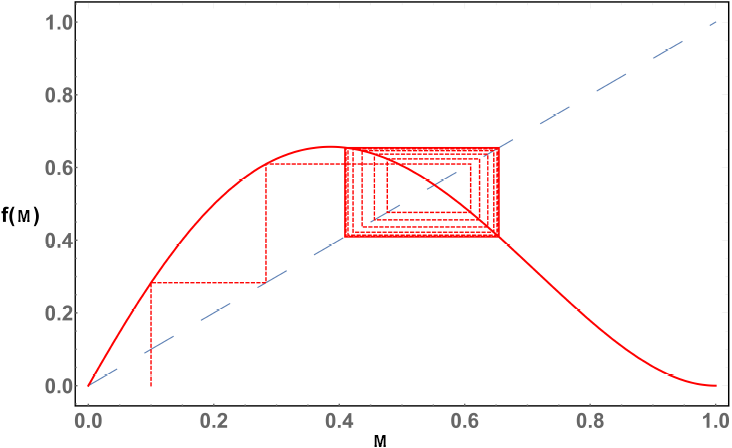}\\
({\bf b}) \\
\includegraphics[width=0.7\linewidth]{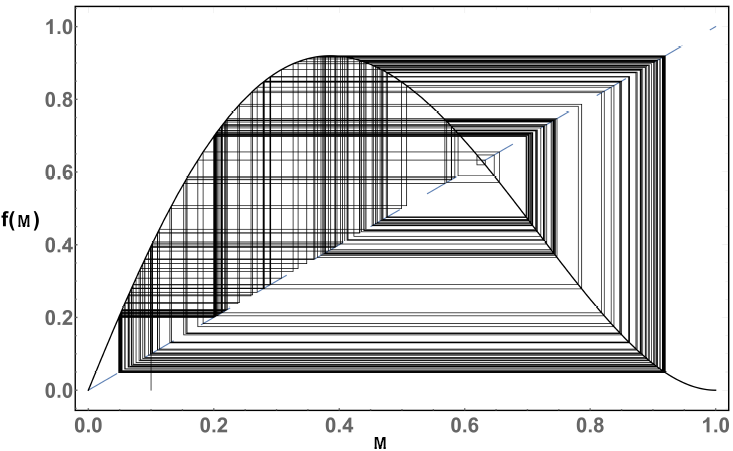}\\
({\bf c}) \\
\end{tabular}
\caption{Cobweb plot of Equation~(\ref{APV}) for three values of the coupling: for $G'= 2.5$ ({\bf a}), the iterative procedure converges as expected and the solution is achieved; for $G'= 2.8$ ({\bf b}), the iterative procedure fails to converge and orbit around the solution of the gap equation; for  $G'= 3.3$ ({\bf c}), any pattern associated with the trajectories of the iterations has disappeared and becomes chaotic.}
\label{fig:orbitchaosPV}
\end{figure}

So far, the iterative procedure has converged as expected.~This means the non-trivial and non-negative solution of the gap equation can  be found through this procedure and is unique. The~first example of the iterative procedure failing to converge is shown in Figure~\ref{fig:iteracionPV}, corresponding  to Equation~(\ref{APV}). The first feature that pops to the eye is the non-converging region of the Figure~\ref{fig:iteracionPV}, similar to the one found in the logistic map \cite{1976Natur.261..459M} . Around $G'\approx2.7$, a second bifurcation appears. The~nature of this flip bifurcation implies that the iterative procedure is bouncing between two values. Moreover,  when $G\approx 3.4$ new bifurcations appear and so on as the coupling grows, until the iteration procedure becomes chaotic. This behavior is better seen in a cobweb plot, as shown in Figure~\ref{fig:orbitchaosPV}. We draw the cobweb plot of three particular values of the coupling constant: For $G'=2.5$ (top), the iteration procedure successfully converges and the solution to the gap equation is achieved; for~$G'=2.8$ (middle), the iterative procedure ceases to converge and each iteration bounces between two values, orbiting around the solution of the gap equation; for $G'=3.3$, we lose any pattern in the iterations and the procedure becomes chaotic. 

\begin{figure}[H]
\centering
\includegraphics[width=0.8\textwidth]{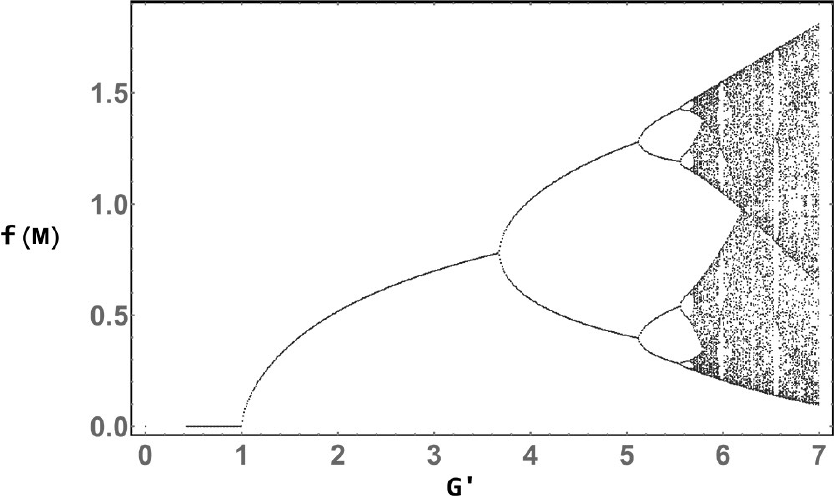}
\caption{{Plot of the last} 70 iterations for each value of the coupling for Equation~(\ref{ATP}). Just as in the Pauli-Villars regularization scheme, we meet with a range of values of the coupling $1<G'< 3.7$ where the iterations converge. After that, bifurcations start to develop and finally, for really strong couplings, $G'> 5.6$ iterations become chaotic.}
\label{fig:iteracionTP}
\end{figure}

Finally, for the PT regularization scheme, Equation~(\ref{ATP}), the  plot of iterations for different couplings can be seen in Figure~\ref{fig:iteracionTP}. In a similar fashion to the PV regularization scheme, firstly we have that only the trivial solution is found though this approach, followed by a behavior consistent with the expected solution for values of $G'>1$, but upon increasing the coupling $G'>3.7$, the iterative procedure  ceases to converge and becomes chaotic. It is worth mentioning that the chaotic  behavior of the PT regularization scheme has been found to hold even in higher dimensions~\cite{Ahmad:2016iez}. Additionally, it should be stressed that even though the iterative procedure does not converge after $G>2.8$ the solution still can be found with others numerical methods, as shown in the same Figure~\ref{fig:iteracionPV}, were we plot the minimum solution of the gap equation Equation~(\ref{APV}) and the iterative process. It should be remarked that the iterative method, if converges, is only capable of finding the solutions that behave as an attractor in the gap equation. This becomes important if the gap equation supports multiple positive solutions. An example of this is Equation~(\ref{APV}) that has two positive solutions for each value of the coupling $G'$. Fortunately, one advantage of solving the gap equation in an iterative manner is that the iterations tend to converge to the smallest of the two solutions, which behaves as an attractor and is the one that minimizes the thermodynamic potential, making it the physically meaningful.

As we have seen, the application of an iterative procedure to solve the gap equation in any regularization scheme may seem straightforward. Nevertheless, we have found that for two of the most popular regularization schemes, such a procedure  fails to converge, albeit for the physically relevant region of coupling constant describing the pion phenomenology in vacuum, the iterations converge smoothly. While it is true that the values of the coupling must be very high for the iterative procedure to fail, it is well known that external factors such as magnetic fields, through the phenomenon of magnetic catalysis, enhance the formation of the chiral condensate  in such a way that the effect of the coupling constant on the gap equation magnifies, thus leaving the possibility that even for small values of the coupling, the iterative procedure cease to converge.
In other regularization schemes, adding the effects of an external magnetic fields is not as straightforward as with the PT regularization, where~we adopt the Schwinger representation of the propagator~\cite{PhysRev.82.664} and use the proper time regularization scheme to get the gap equation in an external magnetic field. Another widely application of the NJL model is the inclusion of the temperature using the Matsubara formalism \cite{kapusta_gale_2006}, and contrary to an~external magnetic field, the thermal bath dilutes the chiral condensate. Therefore, it is worth exploring the behavior of the iteration procedure in a medium, as we do below. 

\section{Gap Equation in A Medium}\label{sec:medium}

As we mentioned before, for its simplicity, the NJL model can be easily extended to account for medium effects as magnetic fields and a thermal bath. The effects of an external magnetic field can be introduced in the model using the Schwinger representation of the fermion propagator, which has the~form
\begin{equation}
\begin{gathered}
iS(p)=\int_0^\infty {\dfrac{ds}{\cos(q_f B s)}e^{is\bigg(p_\parallel^2-p_\perp^2\frac{\tan(q_f B s)}{q_f B s}-m^2+i\epsilon\bigg)}}\times \\
\bigg[(\cos(q_f B s)+\gamma_1\gamma_2\sin(q_f B s))(m+\slashed{p}_\parallel)\bigg]-\dfrac{\slashed{p}_\perp}{\cos(q_f B s)} \;,\label{propM}
\end{gathered}
\end{equation}
where $q_f$ is the absolute value of the quark charge, $2|e|/3$ and $|e|/3$ for the  up and down quarks, respectively, $B$ is the magnetic field strength and we have chosen the homogeneous magnetic field to point in the $\hat{z}$ direction such that $p^\mu_\perp$ and $p^\mu_\parallel$ are defined as
\begin{equation}
\begin{gathered}
p^\mu _\parallel=(p_0,0,0,p_3), \\
p^\mu _\perp=(0,p_1,p_2,0).
\end{gathered}
\end{equation}

Substituting Equation~(\ref{propM}) into Equation~(\ref{condensate}) and taking the trace, we get the chiral condensate in a~homogeneous magnetic field to be given by
\begin{linenomath*}
\begin{equation}
-\langle\bar{\psi}\psi\rangle=4N_{c}m\int\frac{d^{4}p}{(2\pi)^{4}}\int_{1/\Lambda^2}^{\infty}ds\,e^{is\left(p_{\parallel}^{2}-p_{\perp}^{2}\frac{\tan(q_{f}Bs)}{q_{f}Bs}-m^{2}\right)},\label{eq:condensateM}
\end{equation}
\end{linenomath*}
where we are using the simplified notation $\Lambda_{PT}=\Lambda$. To proceed further, we first perform the integration on the perpendicular component of momentum and upon performing a Wick rotation $p_0\rightarrow ip_0$, we get,
\begin{linenomath*}
\begin{equation}
-\langle\bar{\psi}\psi\rangle=\frac{N_{c}mq_{f}B}{\pi}\int\frac{d^{2}p_{\parallel}}{(2\pi)^{2}}\int_{1/\Lambda^2}^{\infty}ds\,\frac{e^{-is\left(p_{\parallel}^{2}+m^{2}\right)}}{\tan(q_{f}Bs)}.
\end{equation}
\end{linenomath*}

Our next goal is to add the effects of temperature. We do so within the Matsubara formalism, through the substitutions 
\begin{linenomath*}
\begin{equation}
\int\frac{dp_{0}}{2\pi} f(p_0)\quad\longrightarrow \quad T\,\sum_{n=-\infty}^{\infty} f(\omega_n),
\end{equation}
\end{linenomath*}
where $\omega_n=2(n+\frac{1}{2})\pi T$ are the Matsubara frequencies. To get the final form of the condensate, we~carry out the remaining momentum integrals and perform a change of variable $s\to-is$, namely,
\begin{linenomath*}
\begin{eqnarray}
-\langle\bar{\psi}\psi\rangle=\frac{N_{f}N_{c}mq_{f}B}{2\pi^{\nicefrac{3}{2}}}\int_{1/\Lambda^2}^{\infty}\frac{ds}{s^{\nicefrac{1}{2}}}\frac{e^{-sm^{2}}}{\tanh(q_{f}Bs)}
  T\sum_{n=-\infty}^{\infty}e^{-s\omega_{n}^{2}}.\label{eq:fullcondensate}
\end{eqnarray}
\end{linenomath*}

With the condensate in Equation~(\ref{eq:fullcondensate}) we can get the gap equation and, in principle, calculate the dynamical mass. Nevertheless, we would like to explicitly separate the contribution of vacuum, which needs regularization. We can identify the sum inside Equation~(\ref{eq:fullcondensate}) with the Jacobi theta function
\begin{linenomath*}
\begin{equation}
\theta_{3}(z,\tau)=1+2\sum_{n=1}^{\infty}q^{n^{2}}\cos(2nz)\;,
\end{equation}
\end{linenomath*}
and $q=e^{i\pi\tau}$. Thus,
\begin{equation}
\begin{array}   {rcl}
\sum\limits_{n=-\infty}^{\infty}e^{-s\omega_{n}^{2}}&=&e^{-\pi^2T^2s}\theta_{3}(i2\pi^2T^2s,i4\pi T^{2}s),\\
&=&(4\pi T^{2}s)^{-\frac{1}{2}}\theta_{3}\left(-\dfrac{\pi}{2},\dfrac{i}{4\pi T^{2}s}\right),\label{eq:inversetheta}
\end{array}
\end{equation}
where we have used the inversion formula~\cite{whittaker_watson_1996}.
\begin{linenomath*}
\begin{equation}
\theta_{3}(\tau,z)=(-i\tau)^{-\frac{1}{2}}e^{-\frac{iz^{2}}{\pi\tau}}\theta_{3}\left(-\frac{z}{\tau},-\frac{1}{\tau}\right).
\end{equation}
\end{linenomath*}

Replacing Equation~(\ref{eq:inversetheta}) into Equation~(\ref{eq:fullcondensate}),  we get the expression for the chiral condensate that includes the thermomagnetic effects of the medium 
\begin{linenomath*}
\begin{eqnarray}
-\langle\bar{\psi}\psi\rangle =\frac{N_{f}N_{c}mq_{f}B}{2\pi^{\nicefrac{3}{2}}}\frac{1}{(4\pi)^{\frac{1}{2}}}\int_{1/\Lambda^2}^\infty\frac{ds}{s}\frac{e^{-sm^{2}}}{\tanh(q_{f}Bs)}
 \theta_{3}\left(-\frac{\pi}{2},\frac{i}{4\pi T^{2}s}\right).\label{eq:condensatewtheta}
\end{eqnarray}
\end{linenomath*}

Finally, the gap equation becomes
\begin{linenomath*}
\begin{eqnarray}
m=m_{q}+GN_{f}N_{c}\frac{mq_{f}B}{2\pi^{2}}\int_{1/\Lambda^2}^{\infty}\frac{ds}{s}\frac{e^{-sm^{2}}}{\tanh(q_{f}Bs)}\theta_{3}\left(-\frac{\pi}{2},\frac{i}{4\pi T^{2}s}\right).\label{eq:ecgapcompleta}
\end{eqnarray}
\end{linenomath*}

Then, to isolate the vacuum contribution to the gap equation, we split the first term of the theta function from the definition
\begin{linenomath*}
\begin{equation}
\theta_{3}\left(-\frac{\pi}{2},\frac{i}{4\pi T^{2}s}\right)=1+2\sum_{n=1}^{\infty}(-1)^{n}e^{-\frac{n^{2}}{4T^{2}s}},
\end{equation}
\end{linenomath*}
and adding and subtracting a factor $\int_{1/\Lambda^2}^{\infty}\frac{ds}{s^{2}}e^{-sm^{2}}$ in  Equation~(\ref{eq:ecgapcompleta}), we are led to


\begin{eqnarray}
\nonumber m & =&m_{q}+GN_{f}N_{c}\frac{m}{2\pi^{2}}\Bigg\{ \int_{1/\Lambda^2}^\infty\dfrac{ds}{s^{2}}e^{-sm^{2}} \\ 
  \nonumber &&\hspace{-8mm}+ \int_{0}^{\infty}\dfrac{ds}{s^{2}}e^{-sm^{2}}\left(\dfrac{q_{f}Bs}{\tanh(q_{f}Bs)}-1\right) \\   
 &&\hspace{-8mm}+ 2q_{f}B\int_{0}^{\infty}\frac{ds}{s}\dfrac{e^{-sm^{2}}}{\tanh(q_{f}Bs)}\sum\limits_{n=1}^{\infty}(-1)^{n}e^{-\frac{n^{2}}{4T^{2}s}}\Bigg\} .\label{eq:gapsplit}
\end{eqnarray}

It should be noted that the integrals in Equation~(\ref{eq:gapsplit}) that involve the magnetic field or the temperature do not require to be regularized. This is so because the divergences in the NJL model appear only in the vacuum term, whereas the medium integrals converge without issues.~Then, Equation~(\ref{eq:gapsplit}) is the expression we were looking for. If we take the limits $B\rightarrow0$ and $T\rightarrow0$, we return to the gap equation in vacuum Equation~(\ref{TP}), as expected. Moreover, from Equation~(\ref{eq:gapsplit}), we may get the respective gap equations for the magnetic case ($T\rightarrow0$), Equation~(\ref{eq:gapmag}), or the thermal case~($B\rightarrow0$), Equation~(\ref{eq:gapter}),
\begin{linenomath*}
\begin{eqnarray}
m=m_{q}+GN_{f}N_{c}\frac{m}{2\pi^{2}}\left\{ \int_{1/\Lambda^2}^{\infty}\frac{ds}{s^{2}}e^{-sm^{2}}+ \int_{0}^{\infty}\frac{ds}{s^{2}}e^{-sm^{2}}\left(\frac{q_{f}Bs}{\tanh(q_{f}Bs)}-1\right)\right\},
 \label{eq:gapmag}
\end{eqnarray}
\end{linenomath*}
\begin{linenomath*}
\begin{eqnarray}
m=m_{q}+GN_{f}N_{c}\frac{m}{2\pi^{2}}\Bigg\{ \int_{1/\Lambda^2}^{\infty}\frac{ds}{s^{2}}e^{-sm^{2}}+ 2\int_{0}^{\infty}ds\frac{e^{-sm^{2}}}{s}\sum_{n=1}^{\infty}(-1)^{n}e^{-\frac{n^{2}}{4T^{2}s}}\Bigg\} .
 \label{eq:gapter}
\end{eqnarray}
\end{linenomath*}

\begin{figure}[H]
\centering
\includegraphics[width=0.8\textwidth]{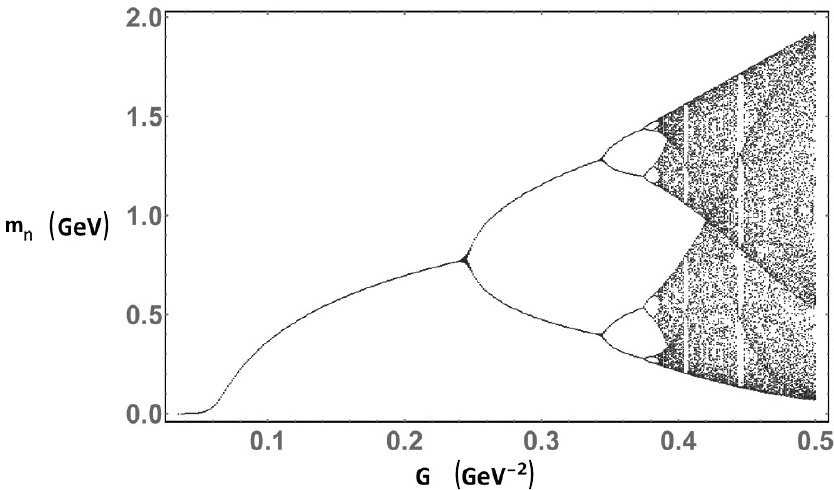}
\caption{\textls[-15]{Plot of the last 70 
iterations performed with Equation~(\ref{eq:gapmag}). The main difference with the vacuum case, Equation~(\ref{ATP}) is the smoothness of the transition between zero mass values and non-zero~values}.}
\label{fig:itermag}
\end{figure}

From here, our next goal is to set up the iteration procedure developed in the previous section to the gap equations Equation~(\ref{eq:gapsplit}), Equation~(\ref{eq:gapmag}) and Equation~(\ref{eq:gapter}), 
but unlike the vacuum case, here~we use  dimensionful quantities.~For the sake of clarity, although we got the gap equation containing both magnetic and thermal contributions, Equation~(\ref{eq:gapsplit}), we  discuss each component separately. Fixing the regularization parameter $\Lambda=1$ and performing the iteration procedure of the above section to  Equation~(\ref{eq:gapmag}), we obtain Figure~\ref{fig:itermag}, with fixed  $eB=0.5\: {\rm GeV}^2$. As we can see, the~behavior of the iterations for different values of the coupling is pretty much the same as that of the gap equation in vacuum, Equation~(\ref{ATP}). Nevertheless, there are a few remarkable differences. First and foremost, unlike the vacuum case, it does not exist  a {\em hard} critical value of the coupling constant $G$ from where the dynamical mass starts being generated. In other words, the derivative of the mass function $M(G)$ obtained is continuous for $G>0$ and up to  the first bifurcation. This can be seen because the transition between a zero value of dynamical mass and a non-zero value of mass is smooth, as a direct consequence of the magnetic field by the magnetic catalysis phenomenon: at a fixed $B$, masses are generated at any weak value of the coupling constant. In fact, this smoothness is responsible for the non-existence of a critical coupling $G_c$ \cite{Martinez:2018snm}.

\begin{figure}[H]
\centering
\includegraphics[width=0.8\textwidth]{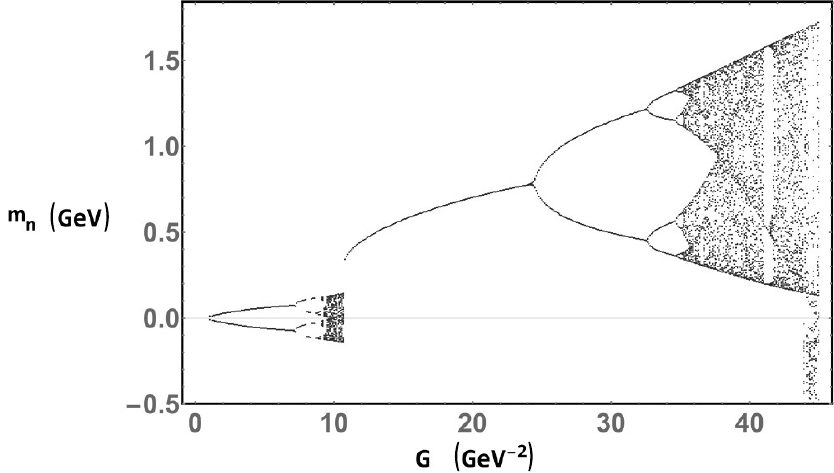}
\caption{Plot of the last 30 of 300 iterations 
performed with Equation~(\ref{eq:gapter}) for $T=0.01\: {\rm GeV}$. As with the vacuum case, Equation~(\ref{ATP}), we find chaos after certain value of the coupling $G\approx25\: {\rm GeV}^{-2}$. On~the other hand, from the beginning and up to $G\approx10\: {\rm GeV}^{-2}$, the iteration procedure fails to converge, bouncing between negative and positive values. What is more, a small region of chaos appears as the coupling approaches $10\:{\rm GeV}^{-2}$.}
\label{fig:iterter}
\end{figure}

\begin{figure}[H]
\centering
\includegraphics[width=.99\textwidth]{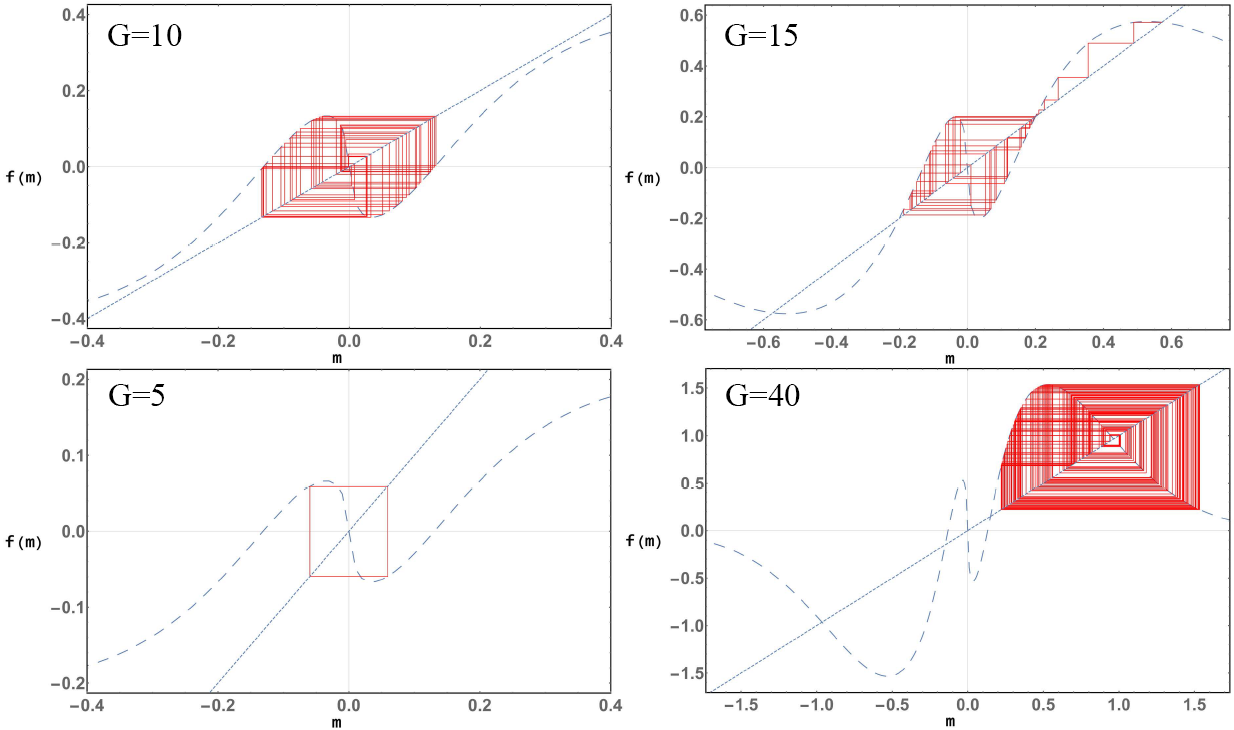}
\caption{Cobweb plot of Equation~(\ref{eq:gapter}) with $T=0.01\: {\rm GeV}$ and various values of the coupling. For~small values of the coupling, the iterations orbit around positive and negative values. Furthermore, for values of coupling close to $G=10\: {\rm GeV}^{-2}$, there appears what seems to be a chaos region. As~the coupling increases, the iterations converge to positive values. Finally, for superstrong couplings $G=40\:{\rm GeV}^{-2}$ the same chaos as in Equation~(\ref{ATP}) appears.}
\label{fig:iterterorbitas}
\end{figure}

Coming up next, in Figure~\ref{fig:iterter} we show the iterations performed with Equation~(\ref{eq:gapter}) for $T=0.1\:{\rm GeV}$. Now the picture changes dramatically. At first glance, the general structure is similar to the free and magnetic cases, but at a closer inspection we find from the beginning and up to $G\approx10\:{\rm GeV}^{-2}$ that the iterations do not converge and, in fact, they bounce taking positive and negatives values. Furthermore, as the coupling approaches $G=10\:{\rm GeV}^{-2}$, it appears what looks like a small version of the chaos found before,  Figures~\ref{fig:iteracionTP} and \ref{fig:itermag}. This can be understood if we take a look at Figure~\ref{fig:iterterorbitas}, where we show the cobweb plot for $G=5\:{\rm GeV}^{-2}$, $G=10\:{\rm GeV}^{-2}$, $G=15\:{\rm GeV}^{-2}$ and $G=40\:{\rm GeV}^{-2}$. On the other side of the spectrum, we confirm the effect of the thermal bath on the condensate, the delay in the breaking of the chiral symmetry, for the iterations to converge the coupling strength must be in the order of $G\approx10\: {\rm GeV}^{-2}$. This is 100 times more than in the magnetic case and around 60 times more than in the vacuum case. One can notice too that the starting point of the first bifurcation, and all the second chaos region in fact, has been delayed due the introduction of the temperature into the system.

\unskip

\begin{figure}[H]
\centering
\includegraphics[width=0.8\textwidth]{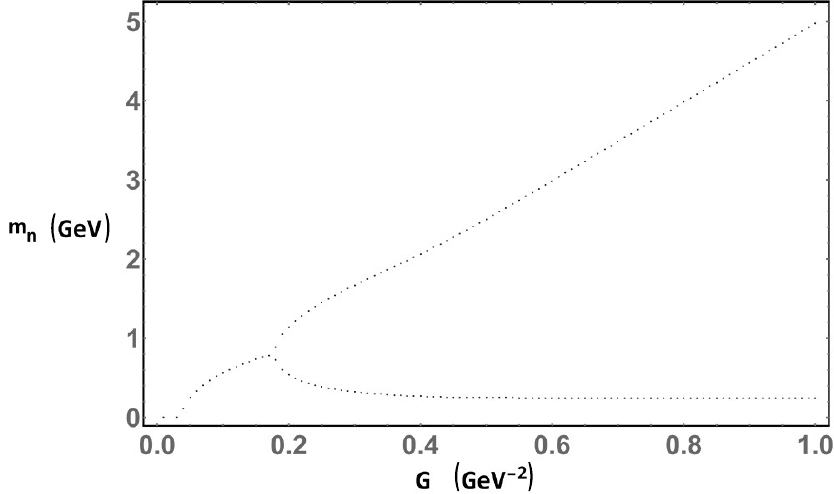}
\caption{Cobweb plot of 
Equation~(\ref{eq:ecgapcompleta}) with $T=0.01\: {\rm GeV}$ and $eB=0.5\:{\rm GeV}^2$. The chaotic part has disappeared. The only remainder is the bifurcation starting at $G\approx0.155\:{\rm GeV}^{-2}$. In this occasion, unlike the magnetic case Figure~\ref{fig:itermag}, we can spot a ``hard'' critical coupling at $G_c\approx0.03\:{\rm GeV}^{-2}$.}
\label{fig:iterTM}
\end{figure}

Finally, Figure~\ref{fig:iterTM} shows the last 70 iterations for $T=0.1\:{\rm GeV}$ and $eB=0.5\:{\rm GeV}^2$. The change is dramatic from the former case. First, the chaotic region is gone, and the only remainder is the sole bifurcation that occurs at $G\approx0.155\:{\rm GeV}^{-2}$. One may be tempted to interpret this bifurcation as two different solutions, but this behavior is solely an effect of the non-convergence of the bifurcations. Another feature is that unlike the magnetic case, the hard breaking of chiral symmetry is present at \mbox{$G_c\approx0.03\:{\rm GeV}^{-2}$}. Physically in the sole magnetic case, $B$ enhances the formation of the chiral condensate making the symmetry breaking {\em soft}, 
where for any value of the coupling, mass is generated, though very small in magnitude. On the other hand, the thermal bath tries to dilute the condensate. In consequence, there appears a gap in the mass spectrum. Put together, the thermal bath forces the apparition of a {\em hard} critical coupling at $G\approx0.155\:{\rm GeV}^{-2}$, meanwhile the magnetic field enhances the chiral condensate. Thus, the coupling necessary for breaking the chiral symmetry goes down~dramatically. 

\section{Conclusions and Final Remarks}\label{sec:final}

\textls[-15]{In this article, we have revisited the commonly used iteration procedure to self-consistently solve the gap equation in the NJL model. Besides the coupling constant, the model has as a~dynamical} parameter a regulator that gives physical reliability to its predictions, namely, by choosing appropriately the coupling and regularization scale, one is able to predict static the properties of pions, for example. Such a regulator can be implemented through several prescriptions. Performing  hard cut-offs on  the three-momentum $3D$ or directly on the four momentum, one observes the need for the coupling to reach a critical value before it is strong enough as to break chiral symmetry. By increasing the coupling constant further, the solution to the gap equation is always found for a sufficiently large number of~iterations.

However, this procedure fails in other regularization schemes such as PV and PT. Though the iteration is stable and converges for small values of the coupling and up to the physically relevant values of this parameter, as it increases to a superstrong regime, the iteration procedure ceases to be stable such that at a second critical point, the iterations bounce between two values separated apart from the actual solution to the gap equation. Increasing further doubles the period and so on until a chaotic behavior unexpectedly emerges. We stress that this behavior indicates the failure of the iteration procedure, not that the gap equation develops new solutions.

Adding the influence of an external magnetic field alone, although similar in shape to that of vacuum, the magnetic catalysis phenomenon taking place in the model makes chiral symmetry to be broken at any weak coupling value. On the other hand, in a thermal bath at temperature $T$, the~iterations make an unexpected behavior bouncing between positive and negative values. Though~the behavior collapses to a stable solution at intermediate values, again, increasing the coupling constant, a chaotic domain in the iterations develop. When the two ingredients are considered at the same time, the chaotic behavior is no longer developed by the iterations in the large coupling regime, though the procedure bounces between two values.

Overall, within this iterative procedure, we observe reasons to take with special care the iteration procedure to obtain the solution to the gap equation in the NJL model at large values of the coupling.

\vspace{6pt} 



\authorcontributions{Conceptualization, methodology, formal analysis, investigation, writing---original draft preparation, writing---review and editing, A.M. and A.R.}

\funding{This research was funded by Consejo Nacional de Ciencia y Tecnolog\'{\i}a (M\'exico) grant number 256494.}

\acknowledgments{We acknowledge Aftab Ahmad and Ricardo Becerril for valuable discussions.}

\conflictsofinterest{The authors declare no conflict of interest.}

\reftitle{References}





\end{document}